\documentclass[english,prl, twocolumn]{revtex4}
\usepackage[T1]{fontenc}
\usepackage[latin9]{inputenc}
\usepackage{babel}
\usepackage{amssymb}
\usepackage{graphicx}
\usepackage{wasysym}
\usepackage{esint}
\usepackage[unicode=true,pdfusetitle,
 bookmarks=true,bookmarksnumbered=false,bookmarksopen=false,
 breaklinks=false,pdfborder={0 0 1},backref=false,colorlinks=false]
 {hyperref}

\makeatletter
\@ifundefined{textcolor}{}
{%
 \definecolor{BLACK}{gray}{0}
 \definecolor{WHITE}{gray}{1}
 \definecolor{RED}{rgb}{1,0,0}
 \definecolor{GREEN}{rgb}{0,1,0}
 \definecolor{BLUE}{rgb}{0,0,1}
 \definecolor{CYAN}{cmyk}{1,0,0,0}
 \definecolor{MAGENTA}{cmyk}{0,1,0,0}
 \definecolor{YELLOW}{cmyk}{0,0,1,0}
}

\makeatother

\begin{document}

\title{Versatile method for renormalized stress-energy computation in black-hole
spacetimes}

\author{Adam Levi and Amos Ori}

\address{Department of physics, Technion-Israel Institute of Technology,\\
Haifa 32000, Israel }
\begin{abstract}
We report here on a new method for calculating the renormalized stress-energy
tensor (RSET) in black-hole (BH) spacetimes, which should also be
applicable to dynamical BHs and to spinning BHs. This new method only
requires the spacetime to admit a single symmetry. So far we developed
three variants of the method, aimed for stationary, spherically symmetric,
or axially symmetric BHs. We used this method to calculate the RSET
of a minimally-coupled massless scalar field in Schwarzschild and
Reissner-Nordstrom backgrounds, for several quantum states. We present
here the results for the RSET in the Schwarzschild case in the Unruh
state (the state describing BH evaporation). The RSET is type I at
weak field, and becomes type IV at $r\lesssim2.78M$. Then we use
the RSET results to explore violation of the weak and null Energy
conditions. We find that both conditions are violated all the way
from $r\simeq4.9M$ to the horizon. We also find that the averaged
weak energy condition is violated by a class of (unstable) circular
timelike geodesics. Most remarkably, the circular null geodesic at
$r=3M$ violates the \emph{averaged null energy condition}.
\end{abstract}
\maketitle
Semiclassical gravity is a theory that describes the interaction of
quantum fields with a classical spacetime metric, and their coupled
evolution. One of the central goals of semiclassical gravity is to
allow detailed understanding of black-hole (BH) evaporation. Since
Hawking's discovery in 1974 that BHs emit quantum radiation \cite{Hawking - Particle creation by black holes}
and evaporate, many efforts have been made to properly formulate and
analyze this dynamical process of semiclassical BH evaporation. This
phenomenon is directly related to a number of profound issues like
the information puzzle and loss of unitarity.

To properly address semiclassical BH evaporation one should (at least
in principle) solve the semiclassical Einstein equation \footnote{Throughout this paper units $c=G=1$ are used, along with signature
($-\mbox{+++}$).} 
\begin{equation}
G_{\alpha\beta}=8\pi\left\langle T_{\alpha\beta}\right\rangle _{ren},\label{eq: Inreo - SC Einstein eq.}
\end{equation}
where $G_{\alpha\beta}$ is the Einstein tensor and $\left\langle T_{\alpha\beta}\right\rangle _{ren}$
is the quantum field's renormalized stress-energy tensor (RSET). Both
sides depend on the evolving spacetime metric $g_{\alpha\beta}(x)$,
which is the unknown in this equation. The RSET, which emerges from
the field's quantum fluctuations, also depends on the type of matter
field as well as on its quantum state. Throughout this paper we shall
consider a minimally-coupled massless scalar field (MCMSF) $\phi(x)$,
satisfying $\square\phi=0$. 

One of the hardest aspects  in dealing with Eq. (\ref{eq: Inreo - SC Einstein eq.})
is the computation of the RSET. In this paper we shall mostly address
this aspect: computation of $\left\langle T_{\alpha\beta}\right\rangle _{ren}$
(and analysis thereof) for a prescribed spacetime metric $g_{\alpha\beta}(x)$.

In principle, this computation involves summation (and integration)
of the contributions to $\left\langle T_{\alpha\beta}\right\rangle $
from the individual field's modes. The naive mode sum is divergent
and requires regularization. This is not surprising, as the naive
expectation value diverges already in flat spacetime. In flat spacetime,
however, one can use the \textit{normal ordering }procedure. Unfortunately,
this simple procedure is not applicable in curved spacetime.

Instead, in curved spacetime one can use the \textit{point-splitting
}regularization which was developed by DeWitt \cite{Dewitt - Dynamical theory of groups and fields}
and later adapted to RSET calculation by Christensen \cite{Christiansen}.
This procedure is based on formally splitting the evaluation point
$x$ into points $x$ and $x'$, and subsequently taking the limit
$x'\to x$ while subtracting a known counter-term.

In its naive form the point splitting procedure is practically inapplicable
to BH backgrounds: Since in this case the field's modes can only be
computed numerically, the naive, direct evaluation of the limit $x'\to x$
becomes impractical. To overcome this problem Candelas, Howard, Anderson
and others \cite{Candelas =000026 Howard - 1984 - phi2 Schwrazschild,Howard - 1984 - Tab Schwarzschild,Anderson - 1990 - phi2 static spherically symmetric,Anderson - 1995 - Tab static spherically symmetric}
developed a version of point-splitting especially adapted to numerical
implementation. In their version, the counter-term subtraction and
the limit $x'\to x$ are analytically translated into certain manipulations
applied to the mode contributions upon summation. This method, however,
is heavily based on (fourth-order)  WKB expansion of the field's modes.
It is therefore inapplicable to dynamical backgrounds (such as evaporating
BHs), because the WKB expansion becomes tremendously difficult in
the time-dependent case. In addition, these methods are in most cases
implemented in the Euclidean sector (to circumvent the turning-point
problem) \textemdash{} which usually does not exist for time-dependent
geometries. The most general case for which this WKB-based method
was applied so far was \cite{Anderson - 1995 - Tab static spherically symmetric}
spherically symmetric static background. \cite{Ottewill-Kerr} Furthermore,
an Unruh state does not exist in the Euclidean sector, so one has
to compute the RSET in a different state (e.g Boulware) in the Euclidean
sector and then compute the (convergent) difference between the two
states in the Lorentzian sector \footnote{For more recent works on regularization (still in the Euclidean sector)
see \cite{Breen and Ottewill - 2012,Breen Hewitt Winstanley and Ottewill - 2015,Ottewill and Taylor - 2010,Ottewill and Taylor - 2011}.}.

Recently we have introduced a new approach to numerically implement
point splitting, which does not rely on the WKB expansion. It can
therefore be implemented directly in the Lorentzian sector. This method
requires the background to admit some symmetry, and the split is made
in the corresponding Killing direction. Our method comes in several
versions, depending on the type of symmetry. We first presented the
$t$-splitting variant, which requires stationarity \cite{Levi =000026 Ori - 2015 - t splitting regularization}.
More recently we have introduced the angular-splitting (or $\theta$-splitting)
variant which requires spherical symmetry \cite{Levi =000026 Ori - 2016 - theta splitting regularization}.
And we have also developed the azimuthal-splitting (or $\varphi$-splitting)
variant which only requires axial symmetry, the details of which will
be presented elsewhere \cite{Preparation}.

In our approach, instead of using WKB, we extract the required short-distance
information directly from Christensen's counter-term. We expand the
latter in the appropriate basis functions (Fourier expansion in $t-t'$
or $\varphi-\varphi'$ variables for $t$- and $\varphi$-splittings
respectively, and Legendre expansion in $\theta-\theta'$ for $\theta$-splitting
\footnote{Here $t,r,\theta,\varphi$ are standard flat-spacetime spherical coordinates
at asymptotically-flat infinity.}). The result of this expansion is then subtracted from the mode contributions,
thereby regularizing their sum.

To introduce and illustrate our approach, in Refs. \cite{Levi =000026 Ori - 2015 - t splitting regularization,Levi =000026 Ori - 2016 - theta splitting regularization}
we employed it to compute $\left\langle \phi^{2}\right\rangle _{ren}$
(rather than the RSET). This is an easier quantity to compute, being
a scalar rather than a tensor, and also being less divergent. This
paper reports the successful calculation of $\left\langle T_{\alpha\beta}\right\rangle _{ren}$
using all three variants: $t$-, $\theta$-, and $\varphi$-splittings. 

We have used our method to compute the RSET in Schwarzschild and Reissner-Nordstrom
backgrounds. Here we focus on the Schwarzschild case, presenting results
for a MCMSF in the Unruh state, using all three variants. Previous
computations of the RSET in the Unruh state were restricted to conformal
fields (scalar \cite{Elster} and electromagnetic \cite{Ottewill}).
Here we compute it for the first time (to our knowledge) for a non-conformal
field. The details of the RSET calculation in the various splittings
will be given elsewhere \cite{Preparation}. 

One of the interesting aspects of quantum-field RSETs is the extent
to which they satisfy or violate various energy conditions. Classical
(minimally coupled \footnote{Non-minimally coupled scalar fields may violate all standard energy
conditions (including ANEC) even classically \cite{Visser-classical}.}) matter fields typically satisfy the \textit{weak energy condition}
(WEC) which states that no observer can measure negative energy density,
namely $T_{\alpha\beta}u^{\alpha}u^{\beta}\geq0$ for any timelike
four-velocity $u^{\alpha}$. Another important energy condition is
the \textit{null energy condition} (NEC), which states that every
null vector $k^{\alpha}$ satisfies $T_{\alpha\beta}k^{\alpha}k^{\beta}\geq0$.
WEC is stronger than NEC and implies it. Such energy conditions are
crucial ingredients in various singularity and horizon theorems \cite{Hawking and Ellis}.
As it turns out, these two purely local conditions are violated by
quantum fields even in flat spacetime \cite{Klinkhammer}. Nevertheless,
 one can also consider averaged conditions. One such important condition
is the \textit{averaged null energy condition} (ANEC) which states
that every null geodesic $x^{\alpha}(\lambda)$ must satisfy
\begin{equation}
\int T_{\alpha\beta}k^{\alpha}k^{\beta}d\lambda\geq0\,,\label{eq:integral}
\end{equation}
where $\lambda$ is an affine parameter and $k^{\alpha}\equiv dx^{\alpha}/d\lambda$.
As it turns out, the ANEC is satisfied \cite{Klinkhammer} in flat
spacetime, and is sufficient for proving several singularity theorems
\cite{Hawking and Ellis} and to enforce topological censorship \cite{Friedman}. 

It has already been shown that the RSET of quantum electromagnetic
field \cite{Roman and Ford - EM field} violates ANEC in the Unruh
state, and the same for conformal scalar field \cite{Visser-U,Visser-B+HH}.
 For a broader discussion on ANEC violations see \cite{Fewster Olum and Pfenning - 2007,Olum}.
Here we use our RSET results to examine the various energy conditions
for a MCMSF  in the Unruh state (for first time as far as we know).

\paragraph{RSET results in the Unruh state.\textendash{}}

\begin{figure}[h]
\centering{}\includegraphics[bb=20bp 180bp 560bp 620bp,clip,scale=0.4]{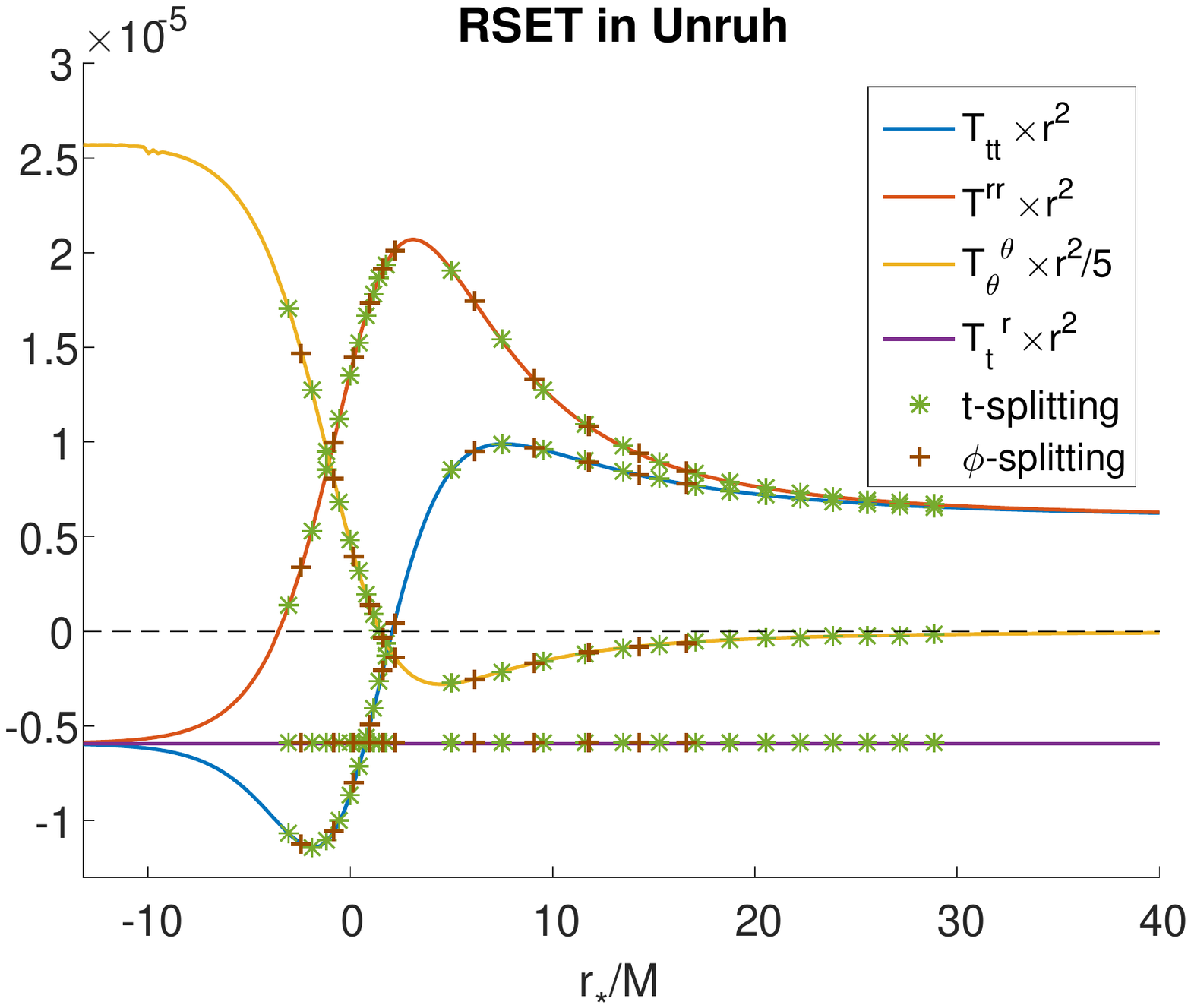}\caption{The various components of the RSET in the Unruh state for a MCMSF
 in Schwarzschild. The curves are the results obtained in $\theta$-splitting,
the asterisks and plus symbols are results obtained using $t$- and
$\varphi$-splittings respectively. Numerical errors start to increase
very close to the horizon (say $r_{*}\apprle-9$), which results in
the tiny wiggles in $T_{\theta}^{\,\,\theta}$. Note that energy-momentum
conservation guarantees that $r^{2}T_{t}^{\,\,r}=const$, yielding
the straight horizontal solid line. \label{fig: Figure 1}}
\end{figure}

The Schwarzschild metric is
\[
ds^{2}=-\left(1-2M/r\right)dt^{2}+\left(1-2M/r\right)^{-1}dr^{2}+r^{2}d\Omega^{2},
\]
where $M$ represents the BH mass and $d\Omega^{2}\equiv d\theta^{2}+\sin^{2}\theta\,d\varphi^{2}$.
 In the background of eternal BH one may consider several distinct
vacuum states. The one which best represents the physics outside a
realistic evaporating BH is the Unruh state \cite{Unruh review}.
We computed the RSET in this quantum state using the three aforementioned
splitting directions. The results of these three computations agree
very well, with typical deviation of order $10^{-3}$ between $\theta$-$t$
splittings, and of order $10^{-2}$ between $\theta$-$\varphi$ splittings.
Figure \ref{fig: Figure 1} displays all (non-vanishing) components
of the RSET \footnote{\label{fn:Units}In all figures, numerical values are expressed in
units of $\hbar/M^{4}$.}, calculated in the three variants, as a function of the tortoise
coordinate $r_{*}=r+2M\ln\left(r/2M-1\right)$.

Since $\varphi$-splitting only assumes axial symmetry, in our spherically-symmetric
background the RSET may in principle be computed at any desired $\theta$
value . The results should of course be independent of $\theta,$
which provides a strong additional consistency check. Here we present
results for $\theta=\pi/2$. We point out, however, that in $\varphi$-splitting
the numerical error typically increases with decreasing $\theta$,
and it becomes harder to use at, say, $\theta\lesssim40^{\circ}$
(which we hope to improve).

Our resultant RSET satisfies energy-momentum conservation, $T_{;\beta}^{\alpha\beta}=0$.
We applied two independent tests to verify this: First, we directly
calculated $T_{;\beta}^{\alpha\beta}$ by applying numerical differentiation
to our numerically-computed RSET. Second, we analytically verified
that all subtraction terms involved in our mode-sum method do satisfy
this conservation law. This is a sufficient check, because the individual
mode contributions trivially satisfy energy-momentum conservation.

One of the conservation equations imposes $r^{2}\left\langle T_{t}^{\,\,r}\right\rangle _{ren}=const$,
yielding the straight solid horizontal line in Fig. \ref{fig: Figure 1}.
Integrating this quantity over the two-sphere gives the total energy
radiated to infinity $L=-4\pi r^{2}\left\langle T_{t}^{\,\,r}\right\rangle _{ren}$.
Our computation yields $L=7.439\cdot10^{-5}\hbar M^{-2},$ which fully
agrees with Elster's result \cite{Elster}.

\paragraph{WEC and NEC analysis.\textendash{}}

The various local energy conditions may conveniently be analyzed through
the eigenvalues $p_{\mu}$ and eigenvectors $V_{(\mu)}^{\alpha}$
($\mu=0,1,2,3$) defined by $T_{\beta}^{\alpha}V_{(\mu)}^{\beta}=p_{\mu}V_{(\mu)}^{\alpha}$.
In spherical symmetry there are always two spacelike eigenvectors
$V_{(2,3)}^{\alpha}$ in the angular directions, with (real) $p_{2}=p_{3}$.
The eigenvalue analysis then reduces to the $2\times2$ matrix $T_{b}^{a}$,
with $a,b=(0,1)$. A direct computation yields the two eigenvalues
$(T_{a}^{a}\pm\sqrt{W}\,)/2$, where $W\equiv2T_{a}^{b}T_{b}^{a}-\left(T_{a}^{a}\right)^{2}$.

In the $W>0$ case $T_{b}^{a}$ admits two real eigenvalues $p_{0}\neq p_{1}$,
associated with two orthonormal eigenvectors $V_{(0,1)}^{\alpha}$
(with $V_{(0)}^{\alpha}$ timelike). Hence altogether $T_{\beta}^{\alpha}$
has four real eigenvalues $p_{\mu}$ with orthonormal eigenvectors
$V_{(\mu)}^{\alpha}$. In the Lorentz frame set by the tetrad $V_{(\mu)}^{\alpha}$
the stress tensor is diagonal, with energy density $\rho\equiv-p_{0}$
and three pressures $p_{i}$. This case is classified in Ref. \cite{Hawking and Ellis}
as type I. NEC is then satisfied iff $\rho+p_{i}\geq0$ for all $i=1,2,3$,
and WEC requires in addition $\rho\geq0$. 

In the $W<0$ case $T_{b}^{a}$ has no real eigenvalues, marking a
type IV \cite{Hawking and Ellis} stress tensor. To analyze NEC we
use double-null coordinates $u,v$ ($g_{uu}=g_{vv}=0$), yielding
$W=4(g^{uv})^{2}\,T_{uu}T_{vv}$. Introducing the two null vectors
$k^{\alpha}=(1,0,0,0)$ and $k^{\alpha}=(0,1,0,0)$, the two projections
$T_{\alpha\beta}k^{\alpha}k^{\beta}$ are just $T_{uu}$ and $T_{vv}$.
It follows that NEC is necessarily violated when $W<0$ \textemdash{}
and so is WEC (a stronger condition).

\begin{figure}
\centering{}\includegraphics[bb=20bp 180bp 560bp 620bp,clip,scale=0.4]{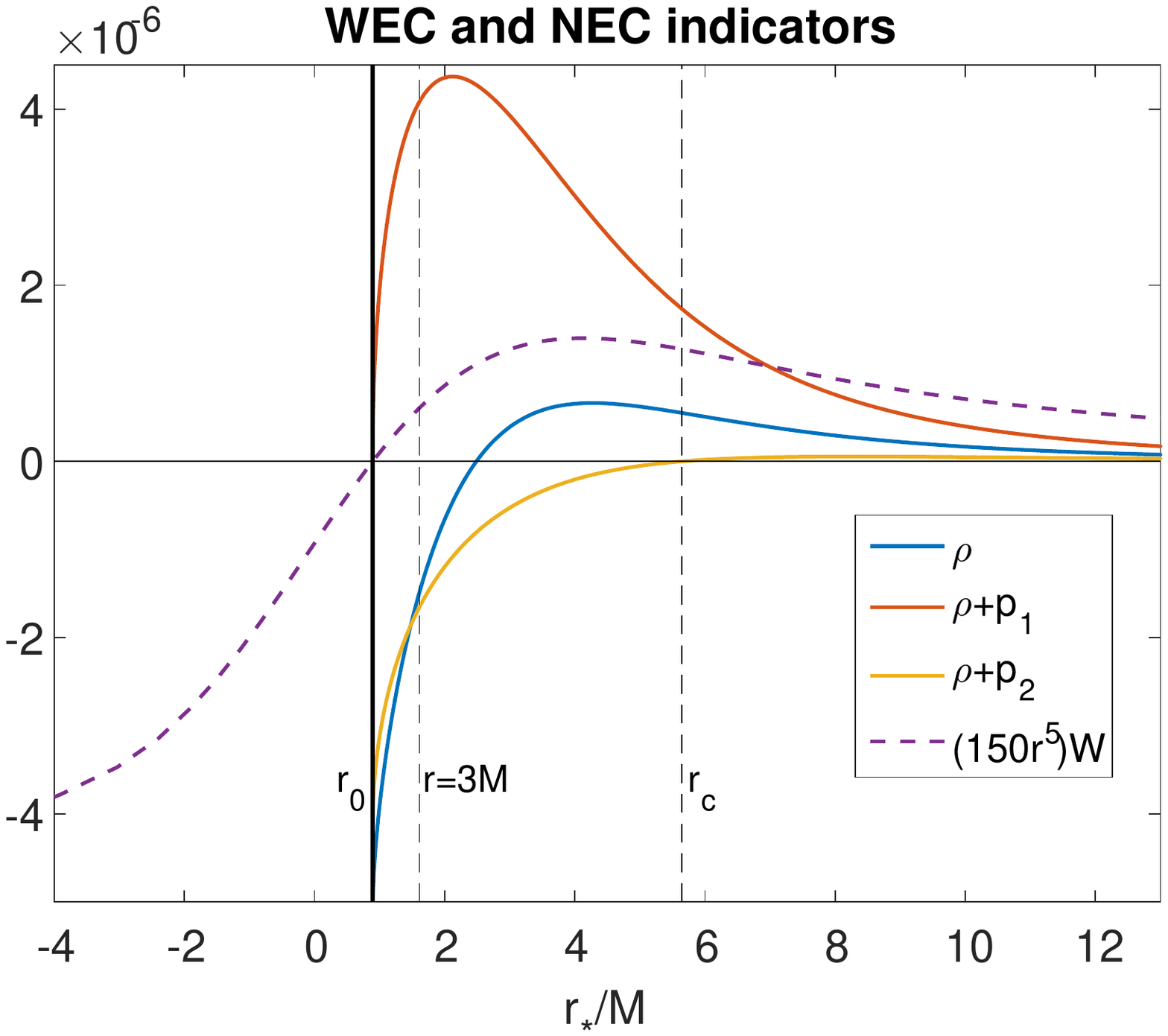}\caption{The various indicators for WEC and NEC, as a function of $r_{*}$.
To the left of the solid vertical line $r=r_{0}\simeq2.78M$ both
WEC and NEC are violated as $W<0$. To the right of $r=r_{0}$, $W$
is positive. NEC then demands $\rho+p_{1},\rho+p_{2}\geq0$, and WEC
also requires $\rho\geq0$. It is clearly seen that $\rho+p_{2}$
turns negative first (with decreasing $r$), hence WEC and NEC break
together. This happens at a surprisingly large radius $r=r_{c}\simeq4.9M$.
WEC and NEC are violated all the way from $r_{c}$ to the horizon.
\label{fig: Figure 2}}
\end{figure}

Figure \ref{fig: Figure 2} displays $W(r)$ (dashed curve) in Schwarzschild's
Unruh state, showing that $W>0$ (i.e. type I) at $r>r_{0}\simeq2.78M$,
and $W<0$ (type IV \footnote{ For earlier observations of type-IV RSET see \cite{Mart=0000EDn-Moruno and Visser - 2013,Mart=0000EDn-Moruno and Visser - 2015}.})
at $2M<r<r_{0}$. It also displays the various WEC/NEC indicators
at $r>r_{0}$ (recall that $p_{3}=p_{2}$ due to spherical symmetry).
One finds that WEC and NEC break together, as the condition $\rho+p_{2}\geq0$
is the first to be violated. This breakdown occurs surprisingly early,
already at $r=r_{c}\simeq4.9M$. It follows that both WEC and NEC
are violated throughout $2M<r<r_{c}$. \footnote{At the transition point $r=r_{0}$ WEC/NEC are violated too, and the
RSET there is type II. }

\paragraph*{Observers along circular geodesics.\textendash{}}

\begin{figure}
\centering{}\includegraphics[bb=20bp 180bp 560bp 620bp,clip,scale=0.4]{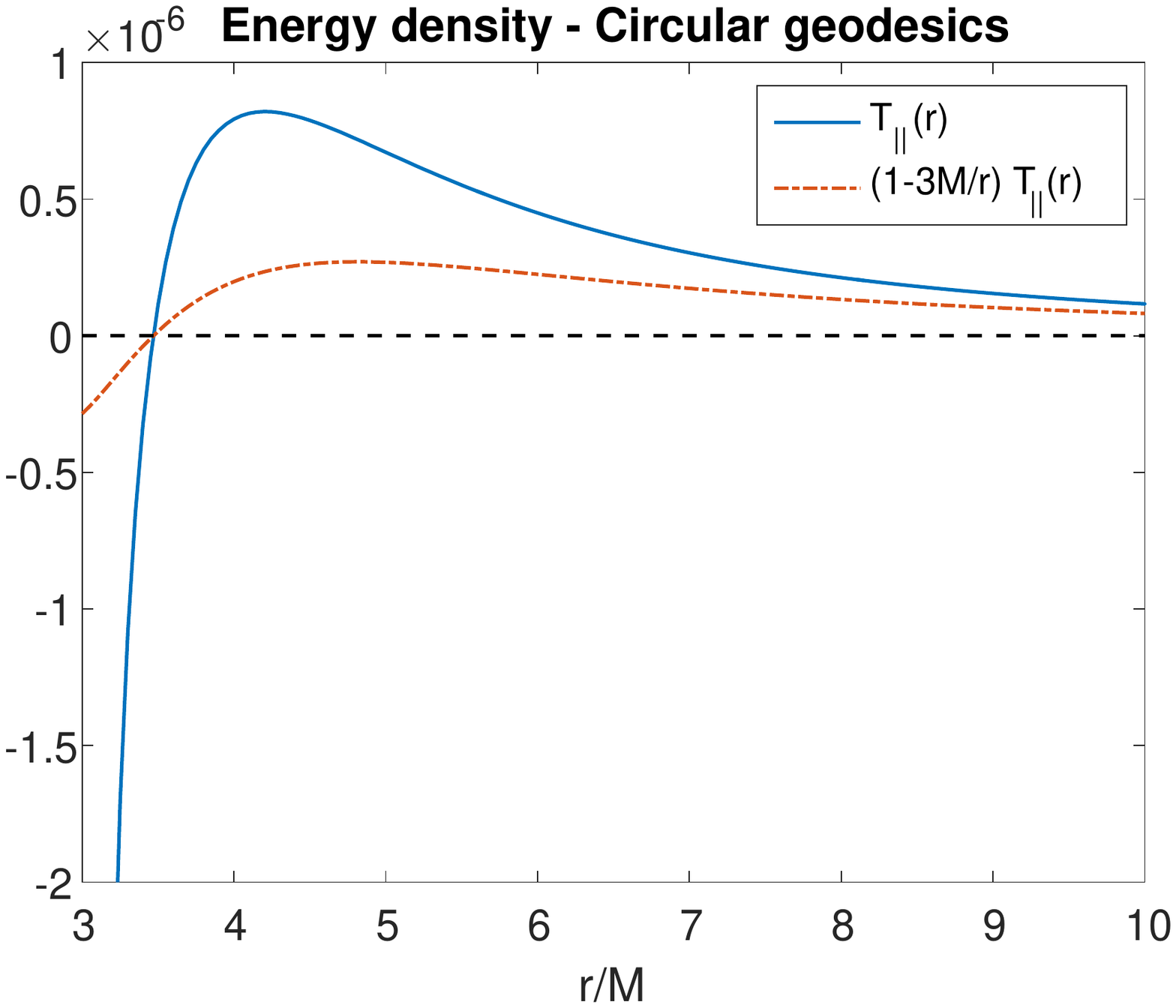}\caption{The projected energy density $T_{\shortparallel}(r)$ along a circular
geodesic at radius $r$. It turns negative at $r\lesssim3.47M$. \label{fig: Figure 3}}
\end{figure}

It is interesting to compute the energy density measured by an observer
moving along a circular geodesic. Assuming an equatorial geodesic,
the projected energy density $T_{\shortparallel}\equiv T_{\alpha\beta}u^{\alpha}u^{\beta}$
is 
\begin{equation}
T_{\shortparallel}(r)=\frac{1}{1-3M/r}\left[\frac{M}{r}\,T_{\phi}^{\,\phi}-\left(1-\frac{2M}{r}\right)\,T_{t}^{\,t}\right].\label{eq:AWEC}
\end{equation}
Figure \ref{fig: Figure 3} plots this quantity as a function of $r$.
It takes its maximum (positive) value at $r\simeq4.2M$ and becomes
negative at $r\lesssim3.47M$. These circular orbits with negative
$T_{\shortparallel}(r)$ are unstable as $r<6M$. Nevertheless this
is an interesting result, as one might hope that although WEC is violated
an averaged version of it might still hold, but these circular geodesics
obviously violate the averaged WEC (AWEC). 

The family of timelike circular geodesics approach a null one as $r\to3M$.
At this limit $T_{\shortparallel}(r)$ diverges, provided that the
term in squared brackets in Eq. (\ref{eq:AWEC}) is $\neq0$ at $r=3M$.
The dashed-dotted curve in Fig. \ref{fig: Figure 3}  indicates that
this quantity is indeed non-vanishing. In fact this limiting quantity
is \emph{negative}, suggesting that the $r=3M$ geodesic should violate
ANEC. We shall now address this issue more directly.

\paragraph*{Violation of the ANEC.\textendash{}}

The most interesting circular orbit is the $r=3M$ null geodesic,
as it is an elegant study case for the ANEC. Its tangent vector is
$k^{\alpha}=a\left(1,0,0,1/\sqrt{27}M\right)$, where $a$ is an arbitrary
constant. The projected energy density is 
\[
T_{\alpha\beta}k^{\alpha}k^{\beta}=\frac{a^{2}}{3}\left[T_{\phi}^{\,\phi}\left(r=3M\right)-T_{t}^{\,t}\left(r=3M\right)\right]\equiv T_{\shortparallel}^{null}\,.
\]
We calculated this quantity and found that 
\[
T_{\shortparallel}^{null}\simeq-2.7\cdot10^{-7}a^{2}\,\hbar M^{-4}\,,
\]
hence this geodesic violates NEC. Furthermore, since $T_{\alpha\beta}k^{\alpha}k^{\beta}$
is constant along the circular orbit, the ANEC is violated too. This
elegant counter-example implies that  the ANEC is not a valid energy
condition in semiclassical curved-spacetime scenarios, and in particular
in BH evaporation. 

\paragraph*{Discussion.\textendash{} }

We demonstrated the usage of our new mode-sum method for numerically
computing the RSET in Schwarzschild background, for a MCMSF in the
Unruh state. So far we developed three variants of this method: $t$-splitting
for stationary backgrounds, $\theta$-splitting for spherically-symmetric
backgrounds, and $\varphi$-splitting for axially-symmetric backgrounds.
Since the Schwarzschild geometry enjoys all these symmetries, we have
been able to compute the RSET in all three variants, and the results
(Fig. \ref{fig: Figure 1}) show nice mutual agreement.  

These three variants complement each other in several ways. First,
in some cases of interest two different variants may be used, which
provides a strong consistency and accuracy check. One notable example
is the Kerr case, in which both $t$- and $\varphi$-splittings are
applicable. Another important example is the time-dependent spacetime
of spherical evaporating BH. Actually this situation, of a spherical
dynamical background, was our main motivation for developing this
mode-sum approach. In this case both $\theta$- and $\varphi$-splittings
should be applicable, which would allow cross-check of the results.
Furthermore, in such a spherically-symmetric situation, the $\varphi$-splitting
computation can be carried at various $\theta$ values, each providing
an independent consistency/accuracy test. 

In addition, there are regions where one of the variants becomes inefficient
or even inapplicable. This typically happens when the coordinate of
symmetry which underlies the splitting direction becomes singular
(or almost singular). For example, in the Schwarzschild case $t$-splitting
deteriorates as one gets close to the horizon, and $\varphi$-splitting
becomes inefficient at small $\theta$ values. Having several different
variants at our disposal allows more effective coverage of the various
spacetime regions. 

The $\varphi$-splitting variant, which was presented here for the
first time, is a powerful method in its own right. The details of
this variant will be presented elsewhere. It should eventually allow
investigation of the dynamical evaporation process of \emph{spinning
BHs} as well. To this end, however, we shall first have to resolve
the difficulties that this variant presently faces at small $\theta$
values. 

The other objective of this paper was the status of various energy
conditions in the spacetime outside a spherical evaporating BH. The
local violation of WEC and NEC by quantum fields is well known; and
indeed we found that both WEC and NEC are violated throughout $2M<r<r_{c}\simeq4.9M$.
Here we showed, however, that \emph{ANEC is violated too} (and the
same for AWEC). In particular, the orbit $r=3M$ violates ANEC. Such
violations were already demonstrated \cite{Visser-U,Roman and Ford - EM field}
for conformally coupled fields, and here we showed it for the first
time (to our knowledge) for a MCMSF.

In fact, this ANEC violation is not limited to the strict $r=3M$
circular geodesic: Consider a ``Zoom-whirl\textquotedbl{} null geodesic
which arrives from infinity and makes a sufficiently large number
of revolutions $N$ around the BH near $r=3M$, before escaping back
to infinity. The ANEC integrand in (\ref{eq:integral}) should be
positive at sufficiently large $r$, but it takes an approximately-constant
negative value ($\approx T_{\shortparallel}^{null}$) near $r=3M$.
As $N$ exceeds some critical value $N_{c}$ (independent of $M$),
the negative contribution will surely dominate, because it is $\propto N$.
Such unbound null geodesics will violate the ANEC too. 

Strictly speaking, this ANEC violation was only shown here for pure
Schwarzschild background. However, such violation must also occur
in \emph{evaporating} BHs. The local backreaction effects on the metric
of an evaporating BH should be $\propto\left\langle T_{\alpha\beta}\right\rangle _{ren}\propto\hbar/M^{4}$.
Consider now a Zoom-whirl null geodesic which makes, say, $N=2N_{c}$
revolutions at $r\cong3M$ around an evaporating BH of (current) mass
$M$. We denote by $\Delta t$ the $t$-interval required for making
these $2N_{c}$ revolutions ($\Delta t\approx12\pi\sqrt{3}\,N_{c}M$).
We can now choose a sufficiently large $M$, such that the BH evaporation
time ($\propto M^{3}$) is $\ggg\Delta t\propto M$. The aforementioned
Zoom-whirl geodesic will hardly be affected by the local backreaction
on the metric, and the same for the ANEC integrand, hence ANEC violation
should occur in this case too.

Note that such ANEC violation should also occur if the BH is spinning,
at least if the spin is not too large, just by continuity. It remains
to investigate whether ANEC violation also occurs in rapidly-spinning
evaporating BHs. Another interesting research direction is using
the RSET results to examine Quantum Inequalities \cite{Ford - 1991,Ford and Roman - 1993,Ford and Roman - 1995,Fewster and Smith - 2008,Olum3}.

It should be noted that a weaker averaged energy condition exists,
the so-called \textit{achronal averaged null energy condition}. This
condition is sufficient \cite{Olum} to prevent certain exotic phenomena.
Kontou and Olum \cite{Olum2} further argued that this energy condition
is guaranteed to hold (for MCMSF) in self-consistent semiclassical
curved spacetimes. The Schwarzschild's $r=3M$ geodesic does not provide
a counter example of course, as it is not achronal. 
\begin{acknowledgments}
\textit{Acknowledgment.\textendash{}} We would like to thank Thomas
Roman and Matt Visser for drawing our attention to Refs. \cite{Roman and Ford - EM field}
and \cite{Visser-U} respectively. We are also grateful to Paul Anderson
for numerous interesting and useful discussions. This research was
supported by the Asher Fund for Space Research at the Technion.
\end{acknowledgments}

\end{document}